# High-ellipticity resonant below-threshold harmonic generation by a helium atom driven by a moderately intense elliptically polarized laser field

M. Yu. Emelin[1] and M. Yu. Ryabikin[1,2]

[1] *Gaponov-Grekhov Institute of Applied Physics of the Russian Academy of Sciences,
46 Ulyanov Street, Nizhny Novgorod 603950, Russia*
[2]*Lobachevsky State University of Nizhny Novgorod,
23 Gagarin Avenue, Nizhny Novgorod 603950, Russia*

Corresponding author: M.Yu. Emelin, emelin@ipfran.ru

The below-threshold harmonic generation in the multiphoton ionization regime for a helium atom driven by the elliptically polarized laser field is studied numerically within the framework of two-dimensional (2D) and three-dimensional (3D) models. It is shown that, regardless of the specific type of atomic potential and the model dimensionality, there is a set of laser frequencies, at which the harmonic generation efficiency increases dramatically with a simultaneous increase of their ellipticity. In these cases, the harmonic ellipticity can exceed the laser ellipticity by 2 or more times in absolute value. It is shown that the efficiency of harmonic generation in this interaction regime depends weakly or almost does not depend on the laser ellipticity. The pathways leading to resonant emission of intense harmonics with increased ellipticities are identified. The enhancement of the ellipticity of these harmonics is explained in terms of relative probabilities of resonant transitions in an elliptically polarized laser field for different values of the magnetic quantum number.

## I. INTRODUCTION

The generation of coherent radiation in the vacuum ultraviolet (VUV) and soft X-ray ranges has long remained a topical area of physics, the interest in which is due to the various applications of such sources for initiating and controlling processes involving electrons within atoms and molecules, as well as for probing various substances [1-3]. One of the most widely used methods for producing coherent VUV and X-ray radiation is the generation of high-order harmonics (HHG) of optical laser radiation in gases [4, 5], typically in the tunneling ionization regime. The spectrum of radiation generated in this process contains a plateau consisting of odd harmonics of the laser field whose energy exceeds the ionization potential of the atom or molecule used for generation [6, 7]. However, from the point of view of both controlling the dynamics of electrons in atoms or molecules and probing certain substances, of great interest is also high-frequency coherent radiation with photon energies below the ionization potential of target atoms or molecules [8-10]. Such radiation can cause not only ionization of the medium, but also effective excitation of atoms or molecules into higher-lying bound states, and often such excitation can proceed in a resonant manner. In addition, coherent high-ellipticity VUV radiation (including that in the 10–30 eV range, close to the ionization thresholds of most molecular systems [11, 12]) is of great interest for a variety of applications; however, the generation of such radiation in a conventional HHG scheme, which involves tunneling ionization of gas particles, is associated with certain difficulties. It is well known that the efficiency of HHG rapidly decreases with increasing ellipticity of the driver laser radiation [13], and the ellipticity of high harmonics, as a rule, does not exceed the laser ellipticity [14]. To solve this problem, various schemes have been proposed [12, 15-21], most of which rely on HHG in two-color fields or on using anisotropic molecules as a nonlinear medium. In both cases, a rather complicated experimental setup is required. Note also that in all the demonstrated schemes, high ellipticity was achieved for a number of moderate-order harmonics with photon energies above the ionization potential, while the generation efficiency was relatively low, as is always the case for harmonics in the plateau region.

In this paper, the process of below-threshold harmonic generation in a helium atom in the multiphoton ionization regime is investigated, with the main attention paid to the possibility of efficient generation of coherent VUV radiation with high ellipticity.

## II. METHODS

This article presents the results of *ab initio* calculations for the helium atom in the single-active-electron approximation. The time-dependent Schrödinger equation (here and below, atomic units are used)

$$i\frac{\partial \Psi}{\partial t} = \frac{1}{2}\left[\vec{p} + \frac{\vec{A}(t)}{c}\right]^2 \Psi + V(\vec{r})\Psi \quad (1)$$

was solved by direct numerical integration using the split-operator method with the fast Fourier transform [22]. Here $\Psi$ is the wave function of the active electron, $\vec{p}$ is the electron momentum operator, $c$ is the speed of light in vacuum, $V(\vec{r})$ is the interaction potential of the electron with the parent ion, and $\vec{A}(t)$ is the vector potential of the laser field. In this work, the laser field was specified as a pulse with a trapezoidal envelope:

$$E_x(t) = \frac{E}{\sqrt{1+\varepsilon_L^2}} f(t)\sin(\omega t),$$
$$E_y\left(t + \frac{T}{4}\right) = \frac{-\varepsilon_L E}{\sqrt{1+\varepsilon_L^2}} f(t)\sin(\omega t), \quad (2)$$

where $E^2$, $\omega$, and $\varepsilon_L$ are the intensity, frequency, and ellipticity of the laser field, respectively, $T = 2\pi/\omega$ is the period of laser field oscillations, and $f(t)$ is a trapezoidal envelope with a flat-top width of 30 field periods:

$$f(t) = \begin{cases} 0, & t < 0, \\ t/(3T), & 0 \leq t \leq 3T, \\ 1, & 3T < t < 33T, \\ 1-(t-33T)/(3T), & 33T \leq t \leq 36T, \\ 0, & t > 36T. \end{cases} \quad (3)$$

The laser intensity was fixed at $10^{14}$ W/cm$^2$, such that the ionization of the atom during the entire pulse did not exceed several percent, even in the case of resonant excitation of intra-atomic transitions. The laser frequency was varied in the range from 0.09 to 0.3 atomic units. The frequency range was chosen so as not to enter the tunnel ionization region, on the one hand, and on the oth-er hand, to include all possible multiphoton resonant transitions from the atomic ground state with a process order equal to 3 or more.

The response of the atom to the laser field was calculated using the Ehrenfest theorem, according to which the second derivative of the atomic dipole moment can be found as follows [23]:

$$\ddot{\vec{d}}(t) = -\frac{d^2}{dt^2}\langle \vec{r} \rangle = \left\langle \Psi(\vec{r},t) \left| \frac{\partial V(\vec{r})}{\partial \vec{r}} + \vec{E}(t) \right| \Psi(\vec{r},t) \right\rangle = \vec{E}(t) + \vec{R}(t), \quad (4)$$

where

$$\vec{R}(t) = \int |\Psi(\vec{r},t)|^2 \frac{\partial V}{\partial \vec{r}} d\vec{r} \quad (5)$$

is the nonlinear part of the atomic response that was later used for the analysis.

The quantity

$$I = \left| F[R_x(t)] \right|^2 + \left| F[R_y(t)] \right|^2, \quad (6)$$

where $F$ denotes the Fourier transform, is the spectral intensity of the nonlinear atomic response, from which the yield of each harmonic can be estimated.

The ellipticity of the generated harmonics, $\varepsilon$, was estimated using the Stokes parameters as follows:

$$\begin{aligned} D_x &= F[R_x(t)], \\ D_y &= F[R_y(t)], \\ S_1 &= |D_x|^2 - |D_y|^2, \\ S_2 &= 2\operatorname{Re}(D_x D_y^*), \\ S_3 &= -2\operatorname{Im}(D_x D_y^*), \\ \varepsilon &= \tan\left[\frac{1}{2}\arctan\left(\frac{S_3}{\sqrt{S_1^2 + S_2^2}}\right)\right], \end{aligned} \quad (7)$$

where Re and Im denote the real and imaginary parts, respectively, and the asterisk denotes complex conjugation.

In the work, two models of the helium atom were used (both in the single-active-electron approximation). Due to the large number of calculations, in order to save computing resources, most of the calculations were made within the framework of a two-dimensional model of the helium atom, represented by the potential

$$V(\vec{r}) = -\frac{1 + (1 + \alpha r)\exp(-\alpha r)}{\sqrt{r^2 + 0.01}} + \frac{0.6 r^6}{r^8 + 0.0001}, \quad (8)$$

where $\alpha = 8.125$. The interaction potential of the active electron with the parent ion was chosen so that the energies of the three lower states of the active electron (1s, 2s, and 2p) in the model atom were close to those in the real helium atom. In potential (8), the energies of the 1s, 2s, and 2p states are -0.9034, -0.1442, and -0.1276 a.u., respectively, which almost perfectly coincide with the corresponding values for the real atom (-0.9036, -0.1459, and -0.1238 a.u., respectively).

To verify the results obtained within the framework of the 2D model, as well as for a more detailed analysis of the problem, calculations were also carried out within the framework of the three-dimensional model of the helium atom, represented by the potential

$$V(\vec{r}) = -\frac{1 + [1 + r\exp(-4r)]\exp(-2r)}{r}, \qquad (9)$$

which was chosen from the same considerations. Within this model, the energies of the 1s, 2s, and 2p states of the active electron are -0.9023, -0.1566, and -0.1287 a.u., respectively, which are also quite close to those for the real helium atom.

## III. RESULTS OF 2D CALCULATIONS

The 2D model of the helium atom made it possible to carry out many series of calculations in a reasonable time with different absolute values of the laser field ellipticity in the range from 0 to 1. Figure 1 shows the dependences of the spectral intensity and ellipticity of the 3rd, 5th, and 7th harmonics on the laser frequency for several (representative) values of its ellipticity.

From the analysis of the data presented in Fig. 1, it follows that for each harmonic of the laser field, there is a certain set of frequencies of this field, at which a local maximum of the spectral intensity of the harmonic is observed, and its ellipticity exceeds $\varepsilon_L$ in absolute value (these points are shown by vertical dotted lines in Fig. 1). The harmonic ellipticity at such points can exceed $\varepsilon_L$ by two or more times. In particular, the absolute value of the ellipticity of the 3rd harmonic can exceed 0.75 at $\varepsilon_L = -0.3$ and be close to 1 at $\varepsilon_L = -0.5$. The absolute value of the ellipticity of the 5th harmonic can exceed 0.6 at $\varepsilon_L = -0.3$ and be about 0.9 at $\varepsilon_L = -0.5$. The absolute value of the ellipticity of the 7th harmonic can be close to 0.9 already at $\varepsilon_L = -0.3$. It can also be noted that the positions of such local maxima do not depend on $\varepsilon_L$, and the energy of the corresponding harmonic does not exceed the ionization potential of the atom, i.e., they all correspond to the below-threshold generation regime. As for the magnitude of the spectral intensity of the harmonic in the selected local maxima, for the 3rd harmonic it depends very weakly on the laser ellipticity; for the 5th harmonic, there are maxima that weakly depend on $\varepsilon_L$, and there are those that decrease more strongly with increasing $|\varepsilon_L|$; for the 7th harmonic, at $|\varepsilon_L| > 0.3$ all the selected local maxima drop off sharply.

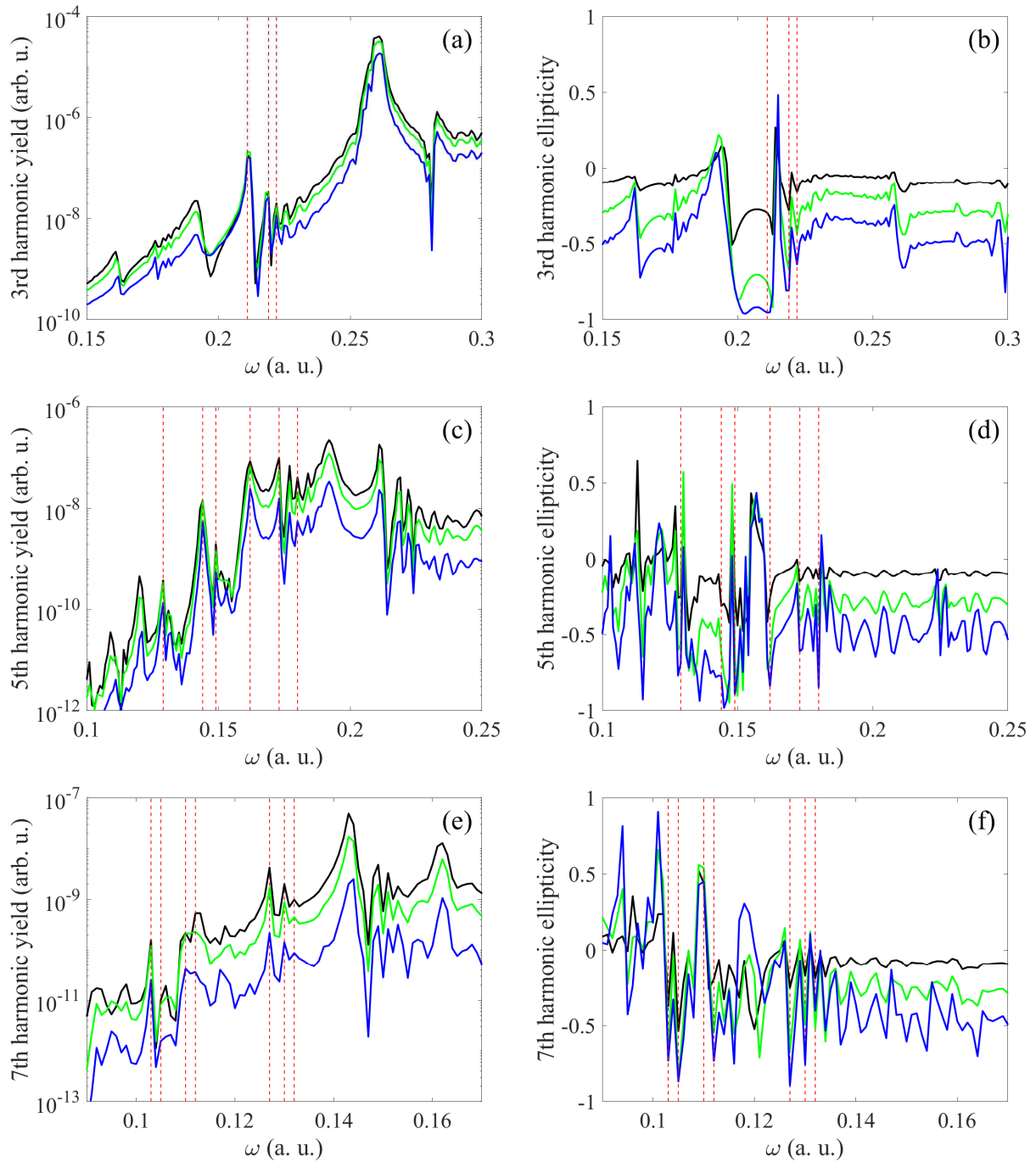

FIG. 1. Dependence of the spectral intensity of the 3rd (a), 5th (c), and 7th (e) harmonics of the laser field and their ellipticity ((b), (d), and (f), respectively) on the frequency of the laser field for different values of its ellipticity ($\varepsilon_L = -0.1$, black line; $\varepsilon_L = -0.3$, green line; and $\varepsilon_L = -0.5$, blue line). The vertical dotted lines on the graphs indicate the points at which a local maximum of the spectral intensity of the corresponding harmonic of the laser field is observed, and its ellipticity exceeds the laser ellipticity in absolute value.

## IV. 3D CALCULATIONS AND RESULTS ANALYSIS

To verify the results obtained within the framework of the 2D model, two series of 3D calculations were made (with $\varepsilon_L = -0.1$ and $\varepsilon_L = -0.3$). Next, we will limit ourselves to the 3rd and 5th harmonics for a more detailed analysis. The results obtained are shown in Fig. 2.

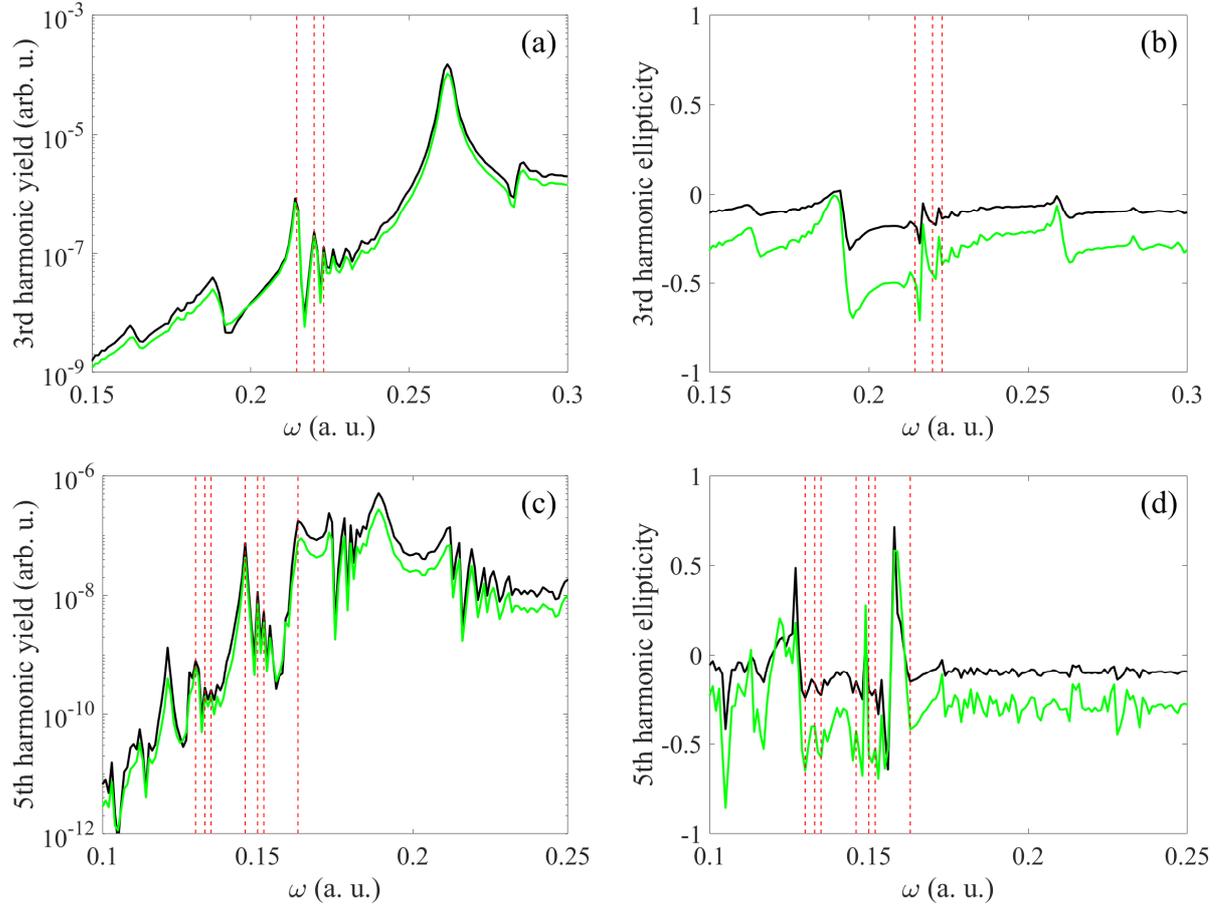

FIG. 2. Dependence of the spectral intensity of the 3rd (a) and 5th (c) harmonics of the laser field and their ellipticity ((b) and (d), respectively) on the frequency of the laser field for different values of its ellipticity ($\varepsilon_L = -0.1$, black line, and $\varepsilon_L = -0.3$, green line). The vertical dotted lines on the graphs indicate the points at which a local maximum of the spectral intensity of the corresponding harmonic of the laser field is observed, and its ellipticity exceeds the laser ellipticity in absolute value.

From the comparison of Figs. 1 and 2 it is evident that the results of the 2D and 3D models are in good qualitative agreement. In the 3D model, there are also points where a local maximum of the spectral intensity of the harmonic is observed, and its ellipticity exceeds in absolute value the laser ellipticity. In particular, the absolute value of the ellipticity of the 3rd harmonic can exceed 0.5, and that of the 5th harmonic can exceed 0.6 at $\varepsilon_L = -0.3$. Just as in the 2D model, the positions of such maxima do not depend on $\varepsilon_L$, and the harmonics are below-threshold. The spectral intensity of the 3rd harmonic at such points is almost independent of $\varepsilon_L$, and that of the 5th harmonic depends weakly.

The behavior of the spectral intensity of harmonics as a function of frequency indicates the resonance nature of the observed peaks. That is, the points marked on the graphs correspond to the multiphoton excitation of one or another intra-atomic transition. However, due to the large number of atomic levels, the close energy of most of them, and the unknown value of the dynamic Stark shift for them, energy considerations alone are insufficient to unambiguously identify a specific intra-atomic transition, the excitation of which corresponds to one or another selected point. It is known *a priori* that before the arrival of the laser pulse the atom is in the ground state. Therefore, each of the maxima on the graphs in Fig. 2 corresponds to a multiphoton excitation of the transition between the atomic ground state and any of the excited states. Thus, our task is to determine only the upper level corresponding to each of the multiphoton resonances. To solve this problem, all (with all possible quantum numbers $l$ and $m$) stationary states in the potential (9) with the principal quantum number n≤4 were found numerically. The quantization axis was chosen to coincide with the direction of the laser pulse wave vector (i.e., the $z$ axis). Due to the elliptical polarization of the laser radiation, this quantization axis is the most natural choice. After that, for several selected points marked with vertical dotted lines in Fig. 2, the projections of the wave function of the active electron onto all the found stationary states during the action of the laser pulse were calculated (the laser ellipticity was fixed at the level of $\varepsilon_L = -0.1$).

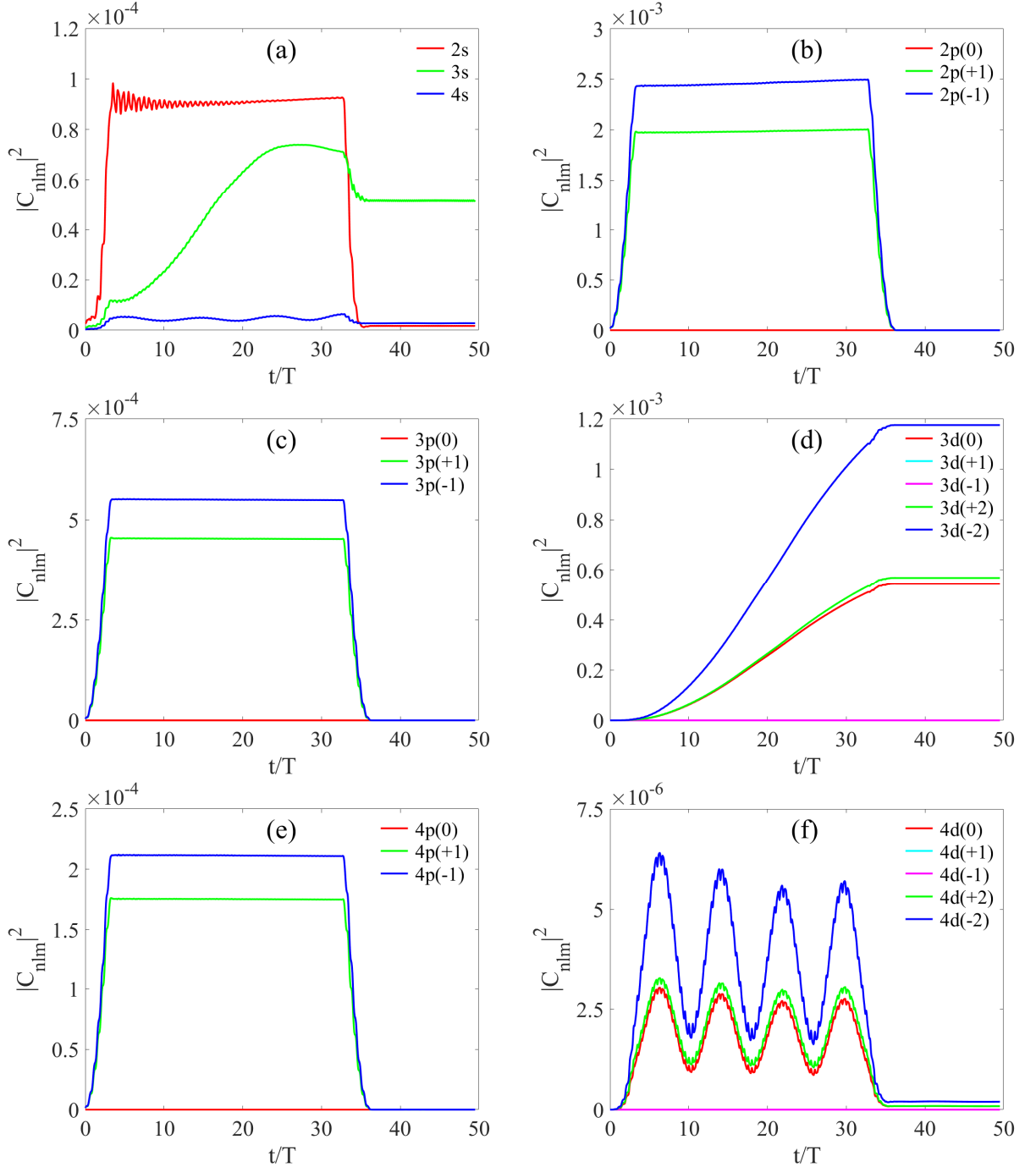

FIG. 3. Projections of the electron wave function of the active electron onto stationary states as a function of time. The frequency of the laser field is 0.214 a. u. Stationary states are specified in the legends, the value of $m$ is given in brackets.

The squares of the moduli of the coefficients corresponding to the projection of the electron wave function onto a particular stationary state, $|C_{nlm}|^2$, for two carrier frequencies of the laser pulse (0.214 and 0.22 a.u.) are shown in Figs. 3 and 4, respectively. To eliminate uninformative oscillations of each curve at the doubled frequency of the laser field, the obtained data were transformed using the moving average method with an averaging interval of one cycle of the laser field. Due to the low degree of excitation and ionization of the atom (the value of the laser field intensity was chosen based on these requirements), the projection of the electron wave function onto the eigenfunction of the ground state is many times greater than the others, and is therefore not shown in the figures (in addition, as noted above, we know *a priori* that it is the lowest level in each of the resonantly excited multiphoton transitions).

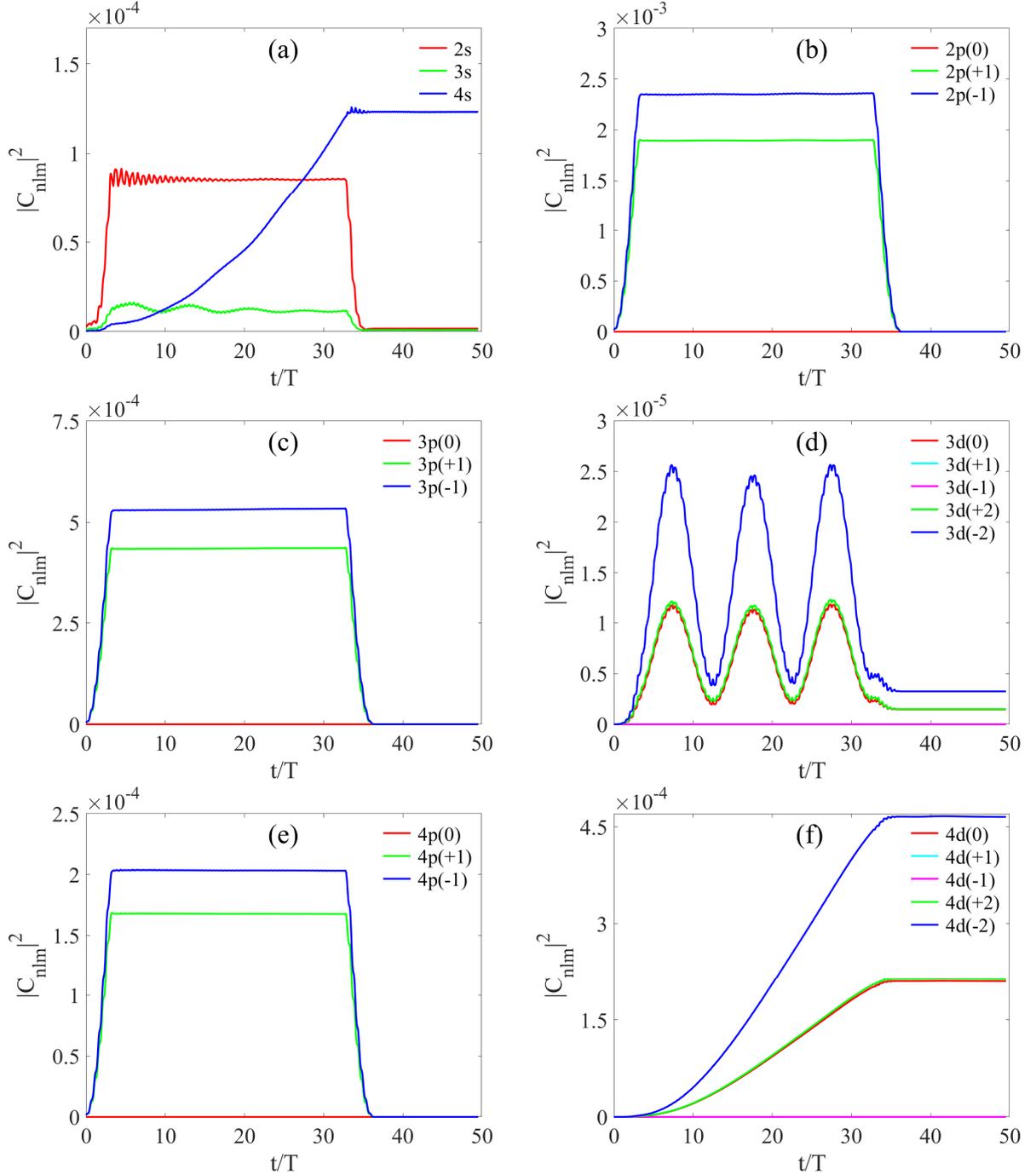

FIG. 4. Same as in Fig. 3, but for a laser frequency of 0.22 a. u.

From the analysis of Fig. 3 it follows that some of the stationary states (namely, 3d and 4d states with |m| = 1, as well as p states with m = 0) are not involved in the process at all, since the projections on them remain zero all the time. Some of the states (namely, p states with |m| = 1) are excited virtually, since the projections on them after the end of the pulse are identically equal to zero. The remaining states (s states and d states with m = 0 and m = 2) are excited really and are significantly involved in the process to a greater or lesser extent. From the comparison of the absolute values of $|C_{nlm}|^2$ and the general form of the curves shown in the figure, it follows that at a laser field frequency of 0.214 a. u., the 1s-3d transition is involved to the greatest extent, and the 1s-3s transition is involved to a lesser extent (with an order of magnitude lower probability), the other states are involved to an even lesser extent. As a result, it can be concluded that at this laser field frequency, the 1s-3d transition is in multiphoton resonance. From energy considerations, it follows that the resonance order is equal to 4.

Carrying out a similar analysis for Fig. 4, we can come to the conclusion that at a laser field frequency of 0.22 a. u., the 1s-4d transition enters the 4-photon resonance. The same calculations and reasoning were done for two of the selected maxima in Fig. 2c. As a result, it was found that at a laser field frequency of 0.146 a. u., the 1s-3d transition enters the 6-photon resonance, and at a frequency of 0.15 a. u., the 6-photon resonance is observed at the 1s-4d transition.

Summarizing the above analysis, we find that effective generation of the 3rd harmonic with a large ellipticity value is observed when the laser field enters a 4-photon resonance with 1s-nd transitions, and for the 5th harmonic, all this occurs when the laser field enters a 6-photon resonance with the same transitions.

## V. MECHANISM FOR EXCITATION OF HARMONICS WITH LARGE ELLIPTICITY

It is obvious that the mechanism underlying all the demonstrated examples of harmonic generation with large ellipticity is the same and does not depend on the harmonic order. Therefore, for simplicity, further reasoning will be given for the 3rd harmonic, and for specificity we will consider the case of multiphoton resonance with the 1s-3d transition.

Since the 1s and 3d states in the case under consideration are related by a four-photon resonance, we are dealing with a four-photon excitation (Fig. 5), and the 3rd harmonic is generated by two-photon de-excitation of an electron from the upper resonant state back to the ground state (with the emission of one quantum at the fundamental frequency of the field and one quantum of the 3rd harmonic, Fig. 6).

For further consideration, it is useful to note that the natural basic states of a photon are circularly polarized states, which are associated with spin angular momenta equal to $\pm\hbar$ (where $\hbar$ is the reduced Planck's constant), and a linearly polarized photon is a photon in a superposition of such states. Accordingly, we will further use the expansions of the field in circularly polarized modes. In our case, the polarization of the laser field is elliptical, which means that we have two populated field modes. In order to find the ratio $X$ of the field amplitude of the less populated circular mode to that of the more populated one under the condition that the ellipticity modulus of the laser field is equal to $|\varepsilon_L|$, we can write the ratio of the semiaxes of the polarization ellipse through the amplitudes of the circular components: $\frac{1-X}{1+X} = |\varepsilon_L|$. Resolving this equation with respect to $X$, we obtain $X = \dfrac{1-|\varepsilon_L|}{1+|\varepsilon_L|}$. Thus, if the number of photons in the more populated mode is taken as $N$, then in the other mode there will be $\left(\dfrac{1-|\varepsilon_L|}{1+|\varepsilon_L|}\right)^2 N$.

From Figs. 3 and 4 we have already clearly seen that in the process under study only d states with $m = 0$ and $|m| = 2$ are populated, which is completely consistent with the law of conservation of angular momentum. Figure 5 schematically shows all possible pathways of 4-photon population of the upper resonant level. The arrows schematically show the photons, and the circles with arrows show the direction of rotation of their electric field.

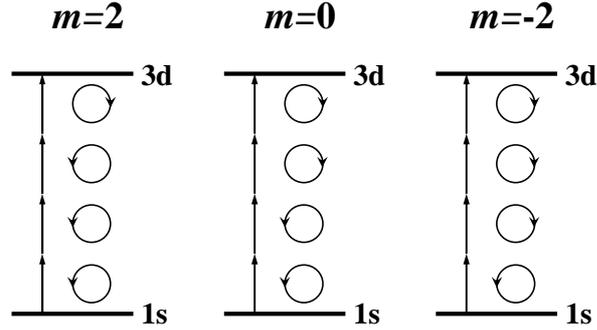

FIG. 5. Pathways of 4-photon population of the upper resonant state.

Figure 6 shows all possible pathways of 2-photon depopulation of the upper resonant state with generation of the 3rd harmonic. It is worth noting that, again according to the law of conservation of angular momentum, states with $|m| = 2$ can be depopulated only with the emission of a photon into the same laser field mode from which a greater number of photons were absorbed during their population, and the photon of the 3rd harmonic must have the same polarization. That is, the depopulation of each of the states with $|m| = 2$ leads to the generation of the 3rd harmonic with circular polarization, but with mutually opposite directions of rotation for $m = 2$ and $m = -2$. As for the state with $m = 0$, its depopulation can occur in two ways. In this case, the polarization of the photon of the 3rd harmonic in each case has a direction of rotation opposite to that of the laser photon.

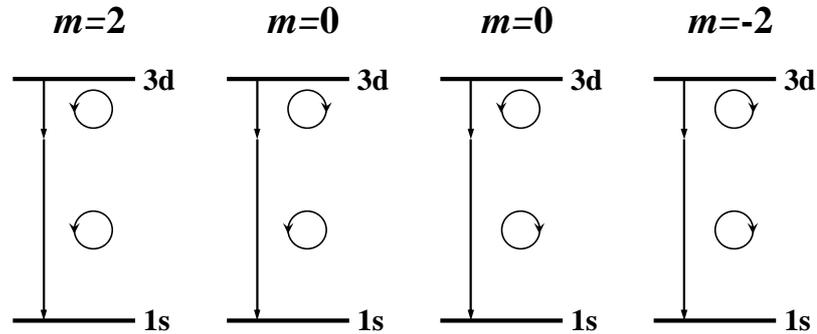

FIG. 6. Pathways of 2-photon depopulation of the upper resonant state with 3rd harmonic generation.

As is known from quantum mechanics, the rate of induced single-photon quantum transitions is linearly proportional to the populations of the initial and final states of the quantum system, the number of photons in the field mode involved in the transition process, and the square of the dipole matrix element of the transition. In the case of a multiphoton process, instead of one matrix element, there will be a complex combination of different matrix elements (a multiphoton matrix element), and instead of the number of photons in one field mode, there will be the product of the populations of all the field modes involved to the degree equal to the number of photons from the given mode involved in the elementary process. Taking into account that the excitation and ionization of the atom under the conditions considered here are small, the population of the lower resonant level (1s) involved in the process can be considered constant and equal to 1. Taking this into account, the expression for the rate of 4-photon excitation of an atom in an elliptically polarized field takes the following form:

$$W_{1s-3d(m)}(t) = \alpha \left(d^{(4)}_{1s-3d(m)}\right)^2 \left(\frac{1-|\varepsilon_L|}{1+|\varepsilon_L|}\right)^{2\gamma} N^4 \left|C_{3d(m)}(t)\right|^2, \qquad (10)$$

where $\alpha$ is a constant, $d^{(4)}_{1s-3d(m)}$ is the 4-photon matrix element, $\gamma$ is the number of photons absorbed from the less populated field mode, and $\left|C_{3d(m)}(t)\right|^2$ is the time-dependent population of the upper resonant state.

In a similar way, one can obtain the following expression for the rate of 2-photon depopulation of the upper resonant state with the generation of the 3rd harmonic of the laser field:

$$W_{3d(m)-1s}(t) = \beta \left(d^{(2)}_{1s-3d(m)}\right)^2 \left(\frac{1-|\varepsilon_L|}{1+|\varepsilon_L|}\right)^{2\eta} N \left(\frac{1-|\varepsilon_{3\omega}|}{1+|\varepsilon_{3\omega}|}\right)^{2\mu} N_{3\omega} \left|C_{3d(m)}(t)\right|^2, \qquad (11)$$

where $\beta$ is a constant, $d^{(2)}_{3d(m)-1s}$ is a 2-photon matrix element, $\eta$ is the number of photons emitted into the less populated mode of the laser field, $\varepsilon_{3\omega}$ is the ellipticity of the field at the 3rd harmonic, $\mu$ is the number of photons of the 3rd harmonic emitted into the less populated mode, and $N_{3\omega}$ is the number of photons in the more populated mode of the 3rd harmonic. Since the field modes at the 3rd harmonic are initially unpopulated in our case, then instead of $N_{3\omega}$ we need to substitute the vacuum value 1/2, and $\varepsilon_{3\omega} = 0$:

$$W_{3d(m)-1s}(t) = \beta' \left(d^{(2)}_{1s-3d(m)}\right)^2 \left(\frac{1-|\varepsilon_L|}{1+|\varepsilon_L|}\right)^{2\eta} N \left|C_{3d(m)}(t)\right|^2. \qquad (12)$$

The population of the upper resonant state in this case (without taking into account ionization) will be determined by

$$\left|C_{3d(m)}(t)\right|^2 = \int_0^t \left(W_{1s-3d(m)}(t') - W_{3d(m)-1s}(t')\right) dt'. \qquad (13)$$

From Eqs. (10), (12), and (13), neglecting the change in the number of photons in laser modes, we can also obtain

$$\frac{d}{dt}\left|C_{3d(m)}(t)\right|^2 = \exp\left[\alpha \left(d^{(4)}_{1s-3d(m)}\right)^2 \left(\frac{1-|\varepsilon_L|}{1+|\varepsilon_L|}\right)^{2\gamma} N^4 t - \beta' \left(d^{(2)}_{1s-3d(m)}\right)^2 \left(\frac{1-|\varepsilon_L|}{1+|\varepsilon_L|}\right)^{2\eta} Nt\right]. \qquad (14)$$

The calculation of multiphoton matrix elements is beyond the scope of this work, so we will limit ourselves to a general analysis of Eqs. (10) and (12) using the diagrams shown in Figs. 5 and 6, respectively.

We begin with a discussion of multiphoton excitation of the atom. It is easy to see that for different $m$ in Eq. (10) only the matrix elements, the number of photons absorbed from a particu-lar mode of the laser field, and the population of the final state differ. It is clearly seen from Fig. 5 that for $\varepsilon_L = -0.1 < 0$ the number of photons absorbed from the less populated mode of the laser field depends on $m$ as follows: $\gamma = 3$ for $m = 2$, $\gamma = 2$ for $m = 0$, and $\gamma = 1$ for $m = -2$. It follows from the latter that the population rates of the upper states with different $m$ differ significantly even for a small value of $\varepsilon_L$ (and given that the populations $|C_{3d(m)}(t)|^2$ are themselves proportional to their population rates, this difference is even greater). Strictly speaking, we cannot compare the population rate of the upper state with $m = 0$ with the others on the basis of such simple reasoning, since the values of the 4-photon matrix elements are unknown to us. However, we note that the squares of these matrix elements for $m = 2$ and $m = -2$ coincide, so the above analysis of the ratio of their population rates is completely correct. From the results of the numerical simulations shown in Fig. 3d, it is clearly seen that the populations $|C_{3d(m)}(t)|^2$ do indeed differ significantly for different values of $m$.

A similar consideration can be made for the 2-photon depopulation of the upper resonant states. In particular, from Eq. (12) and the diagrams shown in Fig. 6 it follows that the depopulation of the upper state with $m = 0$ leads to the generation of the 3rd harmonic of the laser field with ellipticity $\varepsilon_{3\omega} = -\varepsilon_L$. The depopulation of each of the states with $|m| = 2$, as noted above, leads to the generation of the 3rd harmonic of the laser field with circular polarization, but with mutually opposite directions of rotation for $m = 2$ and $m = -2$. Next, we estimate the total ellipticity of the 3rd harmonic generated due to the depopulation of states with $|m| = 2$. Taking into account the equality of the squares of the 2-photon matrix elements for $m = 2$ and $m = -2$, we obtain from Eq. (12) that the ratio of the depopulation rates for these two states is equal to

$$\frac{W_{3d(+2)-1s}(t)}{W_{3d(-2)-1s}(t)} = \left(\frac{1-|\varepsilon_L|}{1+|\varepsilon_L|}\right)^2 \frac{|C_{3d(+2)}(t)|^2}{|C_{3d(-2)}(t)|^2}. \tag{15}$$

For $\varepsilon_L = -0.1$, the first factor in Eq. (15) is $\left(\frac{1-|\varepsilon_L|}{1+|\varepsilon_L|}\right)^2 \approx 0.67$. It follows from Fig. 3d that the ratio of the populations of the two states under discussion remains approximately constant in time and is equal to 0.47. Multiplying the two factors, we have $W_{3d(+2)-1s}(t)/W_{3d(-2)-1s}(t) \approx 0.315$. The depopulation rates of the states are directly proportional to the number of 3rd harmonic photons emitted in the corresponding mode, i.e., $\left(\frac{1-|\varepsilon_{3\omega}|}{1+|\varepsilon_{3\omega}|}\right)^2 \approx 0.315$. Taking square root and resolving this equality with respect to $|\varepsilon_{3\omega}|$, we obtain $|\varepsilon_{3\omega}| \approx 0.281$. From the diagrams presented in Fig. 6, it follows that the ellipticity of this part of the 3rd harmonic radiation has the same sign as $\varepsilon_L$. Thus, for the 3rd harmonic field generated due to the depopulation of states with $|m| = 2$, we finally have $\varepsilon_{3\omega} \approx -0.281$, which significantly exceeds the laser ellipticity. Finally, we should sum up the contributions to the 3rd harmonic caused by the depopulation of states with $m = 0$ and $|m| = 2$, but for this we need to know the 2-photon matrix elements present in Eq. (12), which are unknown to us. However, the total value of the 3rd harmonic ellipticity is also known to us from the numerical simulations presented in Fig. 2b. Namely, for $\varepsilon_L = -0.1$ and the laser field frequency of 0.214 a. u., corresponding to the excitation of the 1s-3d resonance, we have

$\varepsilon_{3\omega} \approx -0.17$. From the above considerations, we see that, although the contribution to the 3rd harmonic signal from the depopulation of the *m* = 0 state (for which $\varepsilon_{3\omega} = -\varepsilon_L$) reduces the total value of the harmonic ellipticity, the value of the latter significantly exceeds $\varepsilon_L$, and this occurs, as has now become clear, due to the high-ellipticity contribution from the states with |*m*| = 2.

Similar reasoning and estimations can be made for the remaining isolated local maxima in Figs. 2a or 2c. It is obvious that the mechanism for generating harmonics with large ellipticity in the discussed regime of laser-atom interaction is based on the multiphoton nature of all intra-atomic transitions occurring here and the significant difference in the rates of population of states with different *m* in cases where the polarization of the laser radiation is different from linear.

## VI. CONCLUSION

In this paper, we have studied the generation of moderate-order harmonics in the regime of excitation of intra-atomic resonances in a multiphoton manner. Based on *ab initio* calculations within two different models of the helium atom, it is demonstrated that during the interaction of an elliptically polarized laser pulse with an atom, there is a set of laser field frequencies at which an increased efficiency of harmonic generation is observed with a simultaneous increase in the absolute value of their ellipticity. It is shown that this set of frequencies corresponds to the multiphoton excitation of certain intra-atomic resonances. In this case, the ellipticity of the harmonics can exceed in the absolute value the laser ellipticity by 2 or more times. It is shown that, in contrast to the case of high harmonics produced in the tunneling regime, the efficiency of harmonic generation in this interaction regime in a number of cases depends weakly or almost does not depend on the laser ellipticity. Based on the analysis of the populations of the atomic stationary states during the action of the laser pulse, it is shown, in particular, that the efficient generation of the 3rd harmonic with a large ellipticity is achieved when the laser field enters the 4-photon resonance between the atomic levels, while for the case of the 5th harmonic this occurs under the conditions of the 6-photon resonance. Based on quantum-mechanical analysis, an explanation is given for the mechanism responsible for the emission of high-ellipticity harmonics. It is shown that this mechanism is associated with the multiphoton nature of the processes occurring and a significant difference in the rates of resonant excitation of upper atomic levels with different values of the magnetic quantum number in an elliptically polarized laser field. Qualitatively, the results obtained and their interpretation do not depend significantly on the specific type of potential and on the dimensionality (2D or 3D) of the model, which indicates the universality of the physical mechanism considered. In general, this generation regime may be of practical interest from the point of view of efficient generation of coherent VUV radiation with high ellipticity and photon energies below the photoionization threshold of atoms or molecules.

## ACKNOWLEDGMENTS

This work was supported by the Russian Science Foundation (project No. 22–12–00389). The authors are grateful to the Joint Supercomputer Center of RAS for the provided supercomputer sources.